\documentstyle[aps]{revtex} 

\author{H. Falomir 
\\ 
Departamento de F\'{\i}sica\\ 
Facultad de Ciencias Exactas,\\  
Universidad Nacional de La Plata, Argentina 
} 
 
\title{Global boundary conditions for the Dirac operator.\thanks{Talk
given at the {\it Trends in Theoretical Physics, CERN - Santiago de Compostela
- La Plata Meeting}, April 27 to May 6, 1997, La Plata, Argentina. } }
 
\date{ May 1 1997 }

\def\dfrac#1#2{{\displaystyle {#1 \over #2}}} 
\def\QATOP#1#2{{#1 \atop #2}} 
\def\QDATOP#1#2{{\displaystyle {#1 \atop #2}}} 
 
\def\dint{\displaystyle \int } 
\def\k{\mbox{\large $\kappa$}} 
\def\ka{\mbox{$\kappa$}} 
\def\nn{\nonumber} 
\def\be{\begin{equation}} 
\def\ee{\end{equation}} 
\def\ker{\rm Ker}

\begin{document} 

\maketitle 
 
\begin{abstract} 

Ellipticity of boundary value problems is caracterized in
terms of the Calderon projector. 
The presence of  topological obstructions for the chiral 
Dirac operator under local boundary conditions 
in even dimension is discussed.
Functional determinants for Dirac operators on 
manifolds with boundary are considered.  

  The functional determinant for a Dirac operator on a bidimensional disk,
in the presence of an Abelian gauge field and  subject to global boundary
conditions of the type introduced by Atiyah-Patodi-Singer, is evaluated.
The relationship with the index theorem is also commented. 
 
\end{abstract}

\section*{Introduction}

The wide application of functional determinants in Quantum and Statistical
Physics is by now a well known fact. In order to evaluate one-loop
effects, one faces to the necessity of defining a regularized determinant
for elliptic differential operators, among which the Dirac first order one
plays a central role. An interesting related problem is the modification
of physical quantities due to the presence of boundaries. The study of
boundary effects has lately received much attention, both in mathematics
and physics, since it is of importance in many different
situations\cite{bordag1,Dowker3:1995,Elizalde1:1996,kirsten-cognola,%
Esposito1:1991,Esposito2:1991,wipf,bolsa-quiral,mitdef}, like index
theorems for manifolds with boundary, effective models for strong
interactions, quantum cosmology and application of QFT to statistical
systems, among others (see \cite{Esposito1:1997} for a recent review). 

In previous work \cite{mochado,bullito}, we studied elliptic Dirac
boundary problems in the case of local boundary conditions. In particular,
we developed for this case a scheme for evaluating determinants from the
knowledge of the associated Green's function, based on Seeley's theory of
complex powers \cite{seeley-cb}. 

Another type of boundary conditions extensively studied in the literature
are global ones, of the type introduced by Atiyah, Patodi and Singer (APS)
\cite{aps} in connection with the index theorem for manifolds with
boundaries (see \cite{egh,Esposito1:1997} for a review.) Other motivation 
for considering these global (or spectral) conditions 
is the
presence of topological obstructions for the chiral Dirac operator under
local boundary conditions (although this restriction no longer holds when
considering the whole Dirac operator \cite{mochado}.)

\section*{Elliptic Boundary Problems and Regularized Determinants}

\subsection*{Elliptic differential operators}

 Let $D$ be a linear differential operator of order $\omega$ in a region
$\Omega$ of ${\rm \bf R}^\nu$,
\be
D=\sum_{|\alpha| \leq \omega} a_\alpha (x) \, (- i \partial_x)^{\alpha}
\ee
(where $\alpha=(\alpha_1, ..., \alpha_\nu)$, $|\alpha|=\alpha_1 +... +
\alpha_\nu$, and the coefficients $a_\alpha (x) \in C^\infty$).
Its {\it symbol} at $x\in \Omega$ is a polynomial in 
$\xi \in {\rm \bf R}^\nu$ of degree $\omega$ defined by
\be
\sigma (D)(x,\xi) = \sum_{|\alpha| \leq \omega} a_\alpha (x) \,  \xi^{\alpha}.
\ee
The {\it principal symbol} of $D$ is the part of $\sigma (D)(x,\xi)$
homogeneous of degree $\omega$ in $\xi$,
\be
\sigma_\omega (D)(x,\xi) = \sum_{|\alpha| = \omega} a_\alpha (x) \,
\xi^{\alpha}.
\ee

An operator $D$ is {\it elliptic} at $x$ if $\sigma_\omega (D)(x,\xi)$ is
invertible $\forall \, \xi \neq 0$. 

If $D$ is elliptic in a compact region $\Omega$ then, for
$ \left| \xi \right|>0$,
\be
\left| \sigma_\omega (D)(x,\xi) \right| \geq constant. \, \left| \xi
\right|^\omega >0, \quad
\forall  \, x\in \Omega,
\ee
since both sides are homogeneous of degree $\omega$, and $a_\alpha (x) \in
C^\infty$.

\bigskip

For example, in ${\rm \bf R}^2$, the operator $D=-i( \partial_1 + i
\partial_2)$ is elliptic, since $\xi_1 + i \xi_2 = 0 \Rightarrow \xi=0$.
The Laplacian, $\nabla=(\partial_1)^2+(\partial_2)^2$ is also elliptic.

\subsection*{Pseudodifferential operators}

Given $f(x) \in {\cal S}( {\rm \bf R}^\nu)$, the Schwartz space, and its
Fourier transform $\hat f (\xi)$, the action of $D$ on $f$ can be
expressed as
\be
D f(x) = \frac 1 {(2\pi)^\nu} \int e^{i x \cdot \xi} 
\sigma (D)(x,\xi) \hat f
(\xi) \, d^\nu \xi.
\label{pseudo}
\ee

\bigskip

More generally, given a smooth function $\sigma (D)(x,\xi)$, with at most
polynomial growth in $\xi$, such that for any $\alpha$ and $\beta$
\be
\left| \partial_{\xi}^{\alpha} \partial_{x}^{\beta}\sigma (D)(x,\xi) \right|
\leq C_{\alpha,\beta} \left(1+ | \xi |\right)^{\omega - \alpha},
\ee
for some constants $C_{\alpha,\beta}$ (with $\omega$ not necessarily
a positive integer), (\ref{pseudo}) defines a {\it
pseudo-differential} operator $D$ of order $\omega$.

A pseudodifferential operator whose symbol decreases faster than any power
of $\xi$ is called {\it infinitely smoothing}. Two pseudodifferential
operators are said to be equivalent if they differ by an infinitely
smoothing operator. This equivalence allows for the introduction of
asymptotic expansions of symbols. 

The basic operation in symbol calculus corresponds to the composition of
operators, and is given by
\be
\left(\sigma_1 \cdot \sigma_2\right)(x,\xi) = 
\sigma_1(x,\xi)e^{-i {\partial^{\leftarrow}\over
{\partial \xi_{\mu}}} {\partial^{\rightarrow}\over{\partial x^{\mu}}}}
\sigma_2(x,\xi) .
\ee


\subsection*{The Calder\'on projector and Elliptic boundary problems}

We will be concerned with boundary value problems 
associated to first order elliptic operators  
\begin{equation} 
\label{OP}D:C^\infty (M,E)\rightarrow C^\infty (M,F),  
\end{equation} 
where $M$ is a bounded closed domain in ${\bf R}^\nu $ with smooth
boundary $\partial M$, and E and $F$ are $k$-dimensional complex vector
bundles over $M.$

In general, such differential operators have 
a closed range of finite codimension, but 
an infinite-dimensional space of solutions,  
\be 
{\rm Ker}(D)=\left\{\varphi(x) / D\varphi(x) =0, \, x\in M\right\}. 
\ee 
Hence, to get a well defined problem, we have to restrict the class of
admissible sections. The natural way of doing this is by imposing boundary
conditions which exclude {\it almost all} solutions of the operator,
leaving only a finite-dimensional kernel. 

In a collar neighborhood of $\partial M$ in $M,$ we will take coordinates
$\bar x=(x,t)$, with $t$ the inward normal coordinate and $x$ local
coordinates for $\partial M$ (that is, $t>0$ for points in $M$ $\setminus
$ $\partial M$ and $t=0$ on $\partial M$ ), and conjugate variables $\bar
\xi =(\xi ,\tau )$. 

\bigskip

One of the most suitable tools for studying boundary problems is the
Calder\'on projector $Q$ \cite{calderon}. For the case we are interested
in, $D$ of order 1 as in (\ref{OP}), $Q$ is a (not necessarily orthogonal)
projection
\be
\label{Q}
Q:L^2(\partial M,E_{/ \partial M})\rightarrow \{T\varphi \ /
\varphi \in {\rm
Ker} (D)\},
\ee
being $T:C^\infty 
(M,E)\rightarrow C^\infty (\partial M,E_{/\partial M})$ the trace map. 

As shown in \cite{calderon}, $Q$ is a zero-th order pseudo differential
operator, and its principal symbol $q(x;\xi ),$ depends only on the
principal symbol of $D,\  \sigma _1(D)$. 

\bigskip 
 
Given any fundamental solution $K(\bar x,\bar y)$ of $D,$ the projector
$Q$ can be constructed in the following way: for $f\in C^\infty (\partial
M,E_{/\partial M})$, one gets $\ \varphi \in \ker (D)$ by means of a
Green formula involving $K(\bar x,\bar y)$, and takes the limit of
$\varphi $ for $\bar x\rightarrow \partial M$. 

Although Q is not uniquely defined, since one can take any fundamental
solution $K$ of $D$ to construct it, the image of $Q$ and its principal
symbol $q(x;\xi )$ are independent of the choice of $K$ \cite{calderon}. 

\bigskip
 
We find it enlightening to compute the principal symbol of the Calder\'on 
projector for the Dirac operator 
\begin{equation}
\label{culo}
D(A)=i\not \! \partial +\not \! \! A=
\sum_{\mu =0}^{\nu -1}\gamma _\mu \left( i \frac \partial {\partial
x_\mu }   + A_\mu\right),
\end{equation}
where $\{A_\mu ,\ \mu =0,...,\nu -1\}$ is the gauge field.
In the present case, $k$ is the dimension of the Dirac spinors in ${\bf
R}^\nu$, $k=2^{[\nu /2]}$.

Let $K(\bar x,\bar y)$ be a fundamental solution of the Dirac operator
$D(A)$ in a neighborhood of the region M$\subset {\bf R}^\nu $, i.e. 
 \begin{equation} 
D^{\dagger }(A)K^{\dagger }(\bar x,\bar y)=\delta (\bar x-\bar y).  
\end{equation} 
We can write  
\begin{equation} 
\label{po}K(\bar x,\bar y)=K_0(\bar x,\bar y)+R(\bar x,\bar y)  
\end{equation} 
where $K_0(\bar x,\bar y)$ is the fundamental solution of 
$i\!\not\!\partial$ vanishing at infinity,
\begin{equation} 
\label{pot}K_0(\bar x,\bar y)=-\ i\ \frac
{\Gamma (\nu /2)}{2\ \pi ^{\nu /2}}%
\ \ \frac{(\bar {\not\! x}-\bar {\not\! y})}
{\vert \bar x-\bar y\vert ^\nu },
\end{equation} 
and $\vert R(\bar x,\bar y)\vert $ is O(1/$\vert \bar x-\bar y\vert ^{\nu
-2})$ for $\left| \bar x-\bar y\right|\sim 0.$

For $f$ a smooth function on $\partial M,$%
\begin{equation} 
\label{poto}Q f(x)=-i\lim \limits_{\bar x\rightarrow \partial 
M}\int_{\partial M}K(\bar x,y)\ \not \! n\ f(y)\ d\sigma _y,  
\end{equation} 
where $\not \!\! n=\sum_l\gamma _l\ n_l,$ and $n=(n_l)$ is the unitary
outward normal vector on $\partial M.$ Note that, if $f=T\varphi $, with
$\varphi \in \ker (D)$, the Green formula yields $Qf=f$, as required. 
 
From (\ref{po}), (\ref{pot}) and (\ref{poto}) one gets  
%
\begin{equation} 
\label{104}  
\begin{array}{c} 
Q f(x)=\frac 12f(x)-i\ P.V. \displaystyle \int_{\partial M}K_0(x,y)
\not \! n\ f(y)\ d\sigma _y\  \\  \\  
-i\ \displaystyle \int_{\partial M}R(x,y)\ \not \! n\ f(y)\ d\sigma _y   .
\end{array} 
\end{equation}
 
To calculate  the principal symbol of $Q$, we write the second term in the
r.h.s. of (\ref{104}) in local coordinates on $\partial M$,
\begin{equation} 
\label{105}\ \ -iP.V.\int_{{\bf R}^{\nu -1}}\frac
{\Gamma (\nu /2)}{2\ \pi 
^{\nu /2}}\ \frac{(x-y)_j}{\vert x-y\vert ^\nu }\ 
\gamma _{j\ }\gamma _n f(y)\ 
dy=\frac 1{2\ }\ \gamma _{j\ }\gamma _n\ {\bf R}_j{\bf (}f)(x),  
\end{equation} 
where ${\bf R}_j(f)$ is the $j$-$th$ Riesz transform of $f$. The symbol of
the operator in (\ref{105}) is (see for example \cite{Stein})
\begin{equation} 
\frac 12i\ \gamma _{j\ }\gamma _n\frac{\xi _j}
{\vert\xi\vert}\ =\frac 12i\  
\frac{\not \! \xi }{\vert \xi \vert }\ \not \! n.  
\end{equation} 
The last term in the r.h.s. of (\ref{104}) is a pseudodifferential
operator of order $\leq -1,$ because of the local behavior of $R(x,y)$,
and then it does not contribute to the calculus of the principal symbol we
are carrying out. Then, coming back to global coordinates, we finally
obtain
 
\begin{equation} 
\label{q}q(x;\xi )=\frac 12(Id_{k\times k}+i\ \frac{\not \! \xi }
{\vert \xi \vert }\ \not \! n).  
\end{equation} 
Note that  
\begin{equation} 
\label{rango}  
\begin{array}{c} 
q(x;\xi )\ q(x;\xi )=q(x;\xi ) \\   
\\  
tr\ q(x;\xi )={k/2},  
\end{array} 
\end{equation} 
and consequently $rank\ q(x;\xi )=k/2.$ 

\bigskip

In particular, for $\nu =2$ and the $\gamma$-matrices given by 
\begin{equation} 
\label{gm}  
\begin{array}{c} 
\gamma _0=\sigma _1=\left(  
\begin{array}{cc} 
0 & 1 \\  
1 & 0  
\end{array} 
\right) ~,\qquad \gamma _1=\sigma _2=\left(  
\begin{array}{cc} 
0 & -i \\  
i & 0  
\end{array} 
\right) ~, \\   
\\  
\qquad \gamma _5=-i\gamma _0\gamma _1=\sigma _3=\left(  
\begin{array}{cc} 
1 & 0 \\  
0 & -1  
\end{array} 
\right) ,  
\end{array} 
\end{equation} 
we obtain  
\begin{equation} 
\label{q2}q(x;\xi )=\left(  
\begin{array}{cc} 
H(\xi ) & 0 \\  
0 & H(-\xi )  
\end{array} 
\right)  
\end{equation} 
$\forall x\in \partial M,$ with $H(\xi )$ the Heaviside function. 
 
\bigskip 
 
According to Calder\'on \cite{calderon}, elliptic boundary conditions can
be defined in terms of $q(x;\xi )$, the principal symbol of the projector
$Q.$
 
{\bf Definition} 1{\sc :} 
 
Let us assume that the $rank$ of $q(x;\xi )$ is a constant $r$ (as is
always the case for $\nu \geq 3$ \cite{calderon}). 
 
A zero-th order pseudo differential operator 
\be
B:[L^2(\partial M,E_{/\partial
M})]\rightarrow [L^2(\partial M,G)],
\ee
with $G$ an $r$ dimensional complex vector bundle over $\partial M,$ gives
rise to an {\it elliptic boundary condition} for a first order operator
$D$ if,
\begin{equation}
\label{erre}
\forall \xi :\vert \xi \vert  \geq 1, \quad
rank(b(x;\xi )\, q(x;\xi ))=rank(q(x;\xi ))=r ,
\end{equation}
where $b(x;\xi )$ coincides with the principal symbol of $B$ for $\vert
\xi \vert \geq 1.$

\bigskip
 
In this case we say that 
\begin{equation} 
\label{BoundaryProblem}\left\{  
\begin{array}{c} 
D\varphi =\chi\ \   
\rm{ in }\ M \\  \\  
BT\varphi =f\ \rm{ at }\ \partial M  
\end{array} 
\right.  
\end{equation} 
is an {\it elliptic boundary problem}, and denote by $D_B$ the closure of
$D$ acting on the sections $\varphi $ $\in C^\infty (M,E)$ satisfying $
B(T\varphi )=0.$
 
 
An elliptic boundary problem as (\ref{BoundaryProblem}) has a solution $
\varphi \in H^1(M,E)$ for any $(\chi ,f)$ in a subspace of $L^2(M,E)\times
H^{1/2}(\partial M,G)$ of finite codimension. Moreover, this solution is
unique up to a finite dimensional kernel \cite{calderon}. In other words,
the operator
\begin{equation} 
(D,BT):H^1(M,E)\rightarrow L^2(M,E)\times H^{1/2}(\partial M,G)  
\end{equation} 
is Fredholm. 

\bigskip
 
For $\nu =2$, Definition 1 not always applies. For instance, for the two
dimensional chiral Euclidean Dirac operator
\begin{equation} 
\label{cdo}D=2i\frac \partial {\partial z^{*}}\ ,  
\end{equation} 
acting on sections with positive chirality and taking values in the
subspace of sections with negative one, it is easy to see from (\ref{q2})
that
\begin{equation} 
\label{qb}q(x;\xi )=H(\xi ).  
\end{equation} 
Then, the $rank$ of $q(x;\xi )$ is not constant. In fact,  
\begin{equation} 
rank\ q(x;\xi )=\left\{ \QDATOP{0\quad \rm{ if }\ \xi <0}
{1\quad \rm{ if }\ \xi >0}\right. .  
\end{equation} 
However, for the (full) two dimensional Euclidean Dirac operator  
\begin{equation} 
\label{fd2}D(A)=\left(  
\begin{array}{cc} 
0 & D^{\dagger } \\  
D & 0  
\end{array} 
\right)  
\end{equation} 
we get from (\ref{rango}) that $rank\ q(x;\xi )=2/2=1$ $\forall \xi \neq
0$, and so Definition 1 does apply.

\subsection*{Local boundary conditions}

When $B$ is a local operator, Definition 1 yields the classical local
elliptic boundary conditions, also called Lopatinsky-Shapiro conditions
(see for instance \cite{hormander}) . 
 
For Euclidean Dirac operators on ${\bf R}^\nu ,$ $E_{/\partial M}=\partial
M\times {\bf C}^k,$ and local boundary conditions arise when the action of
$B $ is given by the multiplication by a $\frac k2\times k$ matrix of
functions defined on $\partial M.$

\bigskip
 
Owing to {\it topological obstructions, }chiral Dirac operators in even
dimensions do not admit{\it \ }local elliptic boundary conditions (see for
example \cite{booss-b}). For instance, in four dimensions, by choosing the
$ \gamma $-matrices at $x=(x_1,x_2,x_3)\in \partial M$ as
\begin{equation} 
\gamma _4=i\left(  
\begin{array}{cc} 
0 & Id_{2\times 2} \\  
-Id_{2\times 2} & 0  
\end{array} 
\right) \ \ \ \rm{and}\qquad \gamma _j=\left(  
\begin{array}{cc} 
0 & \sigma _j \\  
\sigma _j & 0  
\end{array} 
\right) \ \rm{for }\ \ j=1,2,3,  
\end{equation} 
the principal symbol of the Calder\'on projector (\ref{q}) associated to
the full Dirac operator turns out to be
\begin{equation} 
q(x;\xi )=\frac 12\left(  
\begin{array}{cc} 
Id_{2\times 2}+\dfrac{\xi .\sigma }{\vert \xi \vert } & 0 \\  
0 & Id_{2\times 2}-\dfrac{\xi .\sigma }{\vert \xi \vert }  
\end{array} 
\right) .  
\end{equation}
Thus, from the left upper block, one gets for the chiral Dirac operator
\begin{equation} 
q_{ch}(x;\xi )=\frac 12\left(  
\begin{array}{cc} 
1+\dfrac{\xi _3}{\vert \xi \vert } & 
\dfrac{\xi _1-i\xi _2}{\vert \xi \vert } \\  
  &  \\
\dfrac{\xi _1+i\xi _2}{\vert \xi \vert } & 
1-\dfrac{\xi _3}{\vert \xi \vert }  
\end{array} 
\right) .  
\end{equation} 
So $q_{ch}(x;\xi )$ is a hermitian idempotent 2$\times 2$ matrix with $
rank=1.$ If one had a local boundary condition with principal symbol $
b(x)=(\beta _1(x),\beta _2(x))$, according to Definition 1, it should be $
rank(b(x)\ q_{ch}(x;\xi ))=1,\ \forall \xi \neq 0.$ However, it is easy to
see that for
\begin{equation} 
\xi _1=\frac{-2\beta _1\beta _2}{\beta _1^2+\beta _2^2},
\qquad \xi _2=0\quad  
\rm{and\quad }\xi _3=\frac{\beta _2^2-\beta _1^2}
{\beta _1^2+\beta _2^2},  
\end{equation} 
$rank(b(x)\ q_{ch}(x;\xi ))=0.$ 
This is an example of the so called topological obstructions. 
 
Nevertheless, it is easy to see that local boundary conditions can be
defined for the full, either free or coupled, Euclidean Dirac operator 
\be
D(A)=\left( \begin{array}{cc} 0 & D^{\dagger } \\ D & 0 \end{array}
\right) 
\ee 
on $M.$ For instance, we see from (\ref{q2}) and (\ref{erre})
that for $\nu =2$, the operator $B$ defined as
\begin{equation} 
\label{betas}
B\left( \QATOP{f}{g}\right) =
(\beta _1(x),\beta _2(x))\left( \QATOP{f}{g}
\right)  
\end{equation} 
yields a local elliptic boundary condition for every couple of nowhere
vanishing functions $\beta _1(x)$ and $\beta _2(x)$ on $\partial M.$

\subsection*{Global boundary conditions}

A type of non-local boundary conditions to be consider is related to the
ones defined and analyzed by M. Atiyah, V. Patodi and I. Singer in
\cite{aps} for a wide class of first order Dirac-like operators, including
the Euclidean chiral case. Near $\partial M$ such operators can be written
as
\begin{equation}
\varrho \ (\partial _t+{\cal A}),
\end{equation} 
where $\rho:E\rightarrow F$ is an isometric bundle isomorphism,  and 
\be
{\cal A}:L^2(\partial M,E_{/\partial M})\rightarrow 
L^2(\partial M,E_{/\partial
M})
\ee
is self adjoint. The operator $P_{APS}$ defining the boundary condition is
the orthogonal projection onto the closed subspace of $ L^2(\partial
M,E_{/\partial M})$ spanned by the eigenfunctions of ${\cal A}$ associated
to non negative eigenvalues,
\be
P_{APS}=\sum_{\lambda \geq 0} \phi_\lambda 
(\phi_\lambda , \cdot ),\quad{\rm
where\  }{\cal A}\phi_\lambda = \lambda \phi_\lambda.
\ee 
The projector $P_{APS}$ is a zero-th order pseudo differential operator
and its principal symbol coincides with the one of the corresponding
Calder\'on projector \cite{booss-w}. 
 
The problem (\ref{BoundaryProblem})
\begin{equation}
\label{BPro}\left\{
\begin{array}{c}
D\varphi =\chi\ \
\rm{ in }\ M \\  \\
P_{APS}T\varphi =f\ \rm{ at }\ \partial M
\end{array}
\right.
\end{equation}
 with $B=P_{APS}$ has a solution $\varphi \in 
H^1(M,E)$ for any $(\chi ,f)$ with $\chi $ in a finite codimensional 
subspace of $L^2(M,E)$ and $f$ in the intersection of $H^{1/2}(\partial 
M,E_{/\partial M})$ with the image of $P_{APS}$. The solution is unique up 
to a finite dimensional kernel. Note that, since the codimension of $
P_{APS}\ [L^2(\partial M,E_{/\partial M})]$ is not finite, the operator  
\begin{equation} 
(D,P_{APS}T):H^1(M,E)\rightarrow L^2(M,E)\times H^{1/2}(\partial 
M,E_{/\partial M})  
\end{equation} 
is not Fredholm. 

\bigskip

It is to be stressed that, even though $P_{APS}$ has the same principal
symbol as $Q$, their actions are, roughly speaking, {\it opposite}. In
fact, the Calder\'on proyector is related to the problem of the {\it inner
extension} of section over the boundary to global solutions on the
manifold. On the other hand, the action of $P_{APS}$ is related to the
{\it outer extension problem}, in the sense that the solutions of
$D_{P_{APS}}$ admit a square-integrable prolongation on the non-compact
elongation obtained from $M$ by attaching a semi-infinite cylinder
$(-\infty,0] \times \partial M$ to the boundary. 

\bigskip
 
Definition 1 for elliptic boundary conditions does not encompass Atiyah,
Patodi and Singer (APS) conditions since $P_{APS}$ takes values in
$L^2(\partial M,E_{/\partial M})$ instead of $L^2(\partial M,G),$ with $G$
an $r$ dimensional vector bundle ($ r=rank\ q(x;\xi ))$, as required in
that definition. However, it is possible to define elliptic boundary
problems according to Definition 1 by using conditions $\grave a\ la$ APS.
For instance, the following self-adjoint boundary problem for the
two-dimensional full Euclidean Dirac operator is elliptic: 
\begin{equation}
\begin{array}{c}
\left(
\begin{array}{cc}
0 & D^{\dagger } \\
D & 0
\end{array}
\right) \left(
\QDATOP{\varphi_1}{\varphi_2}\right) =\left(
\QDATOP{\chi_1}{\chi_2}\right)
\ \  \rm{ in }\ M, \\  \\
(P_{APS},\varrho (I-P_{APS})\ \varrho ^{*})\left(
\QDATOP{\varphi_1}{\varphi_2}\right) =
h\ \ \rm{ at }\ \partial M,
\end{array}
\end{equation}
 
In fact, as mentioned above, the principal symbol of $P_{APS}$ is equal to
the principal symbol of the Calder\'on projector associated to $D$. So,
from (\ref{qb}) we get
\be
\sigma _0 (P_{APS})(x,\xi)=H(\xi ).
\ee
By taking adjoints we obtain 
\be
\sigma _0(\varrho\, (I-P_{APS})\, \varrho ^{*})=H(-\xi ).
\ee
Then, the principal 
symbol of $B=(P_{APS},\sigma (I-P_{APS})\ \sigma ^{*})$ is 
\be
b(x;\xi )=(H(\xi ),H(-\xi ))
\ee
 and satisfies  
\begin{equation} 
rank(b(x;\xi )\ q(x;\xi ))=rank(q(x;\xi ))\qquad \forall \xi \neq 0.  
\end{equation}

\subsection*{ Functional determinants}

For the case of {\it local boundary conditions} (as in the boundaryless
case), the estimates of Seeley\cite{seeley-cb} allow one to express the
complex powers of $D_B$, $D_B^z$, as an integral operator with continuous
kernel $J_z(x,t;y,s)$ (and, consequently, of trace class) for
$Re(z)<-\nu$. 

As a function of $z$, 
\be
\zeta _{(D_B)}(-z) \equiv Tr(D_B^z)
\ee
can be
extended to a meromorphic function in the whole complex plane {\bf
C}\cite{seeley-trazas}, with only simple poles at $z=j-\nu ,\ j=0,1,2,...$
and vanishing residues for $z=0,1,2,...$

So, in this case, a regularized determinant of $D_B$ can then be defined
as
\begin{equation} 
\label{DR}
Det\ (D_B)=\exp [-\frac d{dz}\ Tr\ (D_B^z)]\vert _{z=0} . 
\end{equation} 
This determinant can also be expressed in terms of the Green's function of
the elliptic boundary value problem, as in \cite{mochado,bullito}. 

\bigskip

But, as far as we know, the construction of complex powers for elliptic
boundary problems with {\it global} boundary conditions is still under
study \cite{seeley-nl}. So, for the global case, one can not use the
previous definition. 

\bigskip

In the following we present the complete evaluation of the determinant of
the Dirac operator on a disk, in the presence of an axially symmetric
Abelian flux and under spectral boundary conditions, in terms of the
corresponding Green's function. 


\section*{Dirac Operator on a disk with global boundary
conditions}

We will evaluate the determinant of the operator $D=\ i \not \!\! \partial
+\not \!\!\! A$ acting on functions defined on a two dimensional disk of
radius $R,$ under APS boundary conditions. 

We consider an Abelian gauge field in the Lorentz gauge, $A_\mu =\epsilon
_{\mu \nu }\ \partial_\nu \phi$ $\ (\epsilon _{01}=-\epsilon _{10}=1)$,
with $\phi $ a smooth bounded function $\phi =\phi (r)$; then
\be
A_r=0, \qquad A_\theta(r)=-\partial_r\phi
(r)=-\phi^{^{\prime }}(r).
\ee
The flux through the disk is 
\begin{equation}
\label{3.2}
{\k } =\dfrac{\Phi}{2\pi}=\dfrac{1}{2\pi}\oint_{r=R}\ A_\theta
\ R\ d\theta =- R\phi^{^{\prime }}(R).
\end{equation}

With the conventions for the $\gamma$-matrices stated above, the full
Dirac operator can be written as
\begin{equation}
 D=e^{-\gamma _5\phi
(r)\ }i\not \! \partial \ e^{-\gamma _5\phi (r)}
=\left(
\begin{array}{cc} 0 & \varrho
^{-1}(\partial _r+{\cal A}) \\ \varrho \ (-\partial _r+{\cal A}) & 0
\end{array} \right),
\label{op}
\end{equation}
where 
\begin{equation}
\varrho =-\ i\ e^{i\theta },
\end{equation}
and
\begin{equation}
{\cal A}(r)=-\frac ir\ \partial _\theta +\ \partial
_r\phi (r).
\label{beder}
\end{equation}
At the boundary, the eigenvectors and eigenvalues of the self adjoint
operator ${\cal A}(R)$ are given by
\be
{\cal A}(R)\,e^{in\theta } =a_n\, e^{in\theta }, 
\quad {\rm with\ }a_n = \frac
1R(n-{\k}), \   n \in {\rm \bf Z}.
\ee
We take the radial variable to be conveniently adimensionalized throught
multiplication by a fixed constant with dimensions of mass. 

Let $k$ be the integer such that $k<{\k } \leq k+1$.
We will consider the action of the differential operator $D$ on the space
of functions satisfying homogeneous global boundary conditions characterized by
\begin{equation}
\label{apsbc}\left(
\begin{array}{cc}
{\cal P}_{\geq} & \varrho\, (1-{\cal P}_{\geq})\, \varrho ^{*}
\end{array}
\right) \left( \QATOP{\varphi (R,\theta )}{\chi (R,\theta )}\right) =0,
\end{equation}
where 
\begin{equation}
{\cal P}_{\geq}=\frac 1{2\pi }\sum_{n\geq k+1}e^{in\theta }\left(
e^{in\theta
},\ \cdot \ \right),
\end{equation}
\begin{equation}
\varrho\, (1-{\cal P}_{\geq})\, \varrho^{*}=\varrho \, {\cal P}_{<}\,
\varrho^{*}
=\frac
1{2\pi }\sum_{n\leq k+1}e^{in\theta }\left( e^{in\theta
},\ \cdot \ \right) ={\cal P}_{\leq} \, .
\end{equation}
Notice that the operator so defined, which we call 
$\left( D\right) _{{\ka} }$, turns out to be self adjoint. 

\bigskip

Our aim is to compute the quotient of the determinants of the operators $
\left( D\right) _{{\ka} }$ and $(i\not \! \partial )_{{\ka }=0}.$ Since
the global boundary condition in Eq. (\ref{apsbc}) is not a continuous
function of the flux $\Phi$, we will proceed in two steps: 
\begin{equation}
\left( D\right) _{{\ka } }\rightarrow (i\not \! \partial
)_{{\ka }}\rightarrow (i\not \! \partial  )_{{\ka }=0} .
\end{equation}
In the first step, where there is no change of boundary conditions, we can
grow the gauge field by varying $\alpha$ from $0$ to $1$ in
\begin{equation}
\label{opalf}D_\alpha =i\not \! \partial +\alpha \not \! \!
A=e^{-\alpha \gamma
_5\phi (r)\;\ }i\not \! \partial \ e^{-\alpha \gamma _5\phi (r)},
\end{equation}
thus going smoothly from the free to the full Dirac operator. The explicit
knowledge of the Green's function will allow us to perform the calculation
of this step, where we will use a gauge invariant point splitting
regularization of the $\alpha$-derivative of the determinant. The second
step will be achieved by using a $\zeta$-function regularization, after
explicitly computing the spectra.

There is an additional complication, since these global boundary
conditions give rise to the presence of $\vert k+1 \vert$ linearly
independent zero modes. For $k>0$, these normalized eigenvectors are given
by
\begin{equation}
\dfrac{\ e^{\alpha \phi (r)}}{\sqrt{2\pi \ q_n(R;\alpha)}}\ \left(
\QDATOP{X^n}{0}
\right) ,{\rm \ \   with }\ 0\leq n\leq k,
\label{ceros}
\end{equation}
where $X=x_0 + i x_1=r e^{i \theta}$, and the normalization factors are 
\begin{equation}
q_n(u;\alpha)=\dint_0^u\ e^{2\alpha \phi (r)}\ r^{2n+1}\ dr .
\end{equation}
For $k<1$ we get similar expressions, with the opposite chirality.  Notice
that, for $k=-1$ (in particular, when $\Phi =0)$, there is no zero mode. 

\bigskip

For simplicity, in the following we will consider only the case $k\geq
-1$. The kernel of the orthogonal projector on Ker$(D_\alpha )_{{\ka}}$,
$P_\alpha $ is given by

\begin{equation}
P_\alpha (z,w)=\sum\limits_{n=0}^k\dfrac{\ e^{\alpha
[\phi (z)+\phi (w)]}}{
2\pi \ q_n(R;\alpha)}\left(
\begin{array}{cc}
(ZW^{*})^n & 0 \\
0 & 0
\end{array}
\right) .
\end{equation}

\bigskip

Now, since  $(D_\alpha +P_\alpha )_{{\ka }}$ is  invertible, we can define
\begin{equation}
Det^{^{\prime }}(D_\alpha )_{{\ka }}\equiv Det(D_\alpha +P_\alpha
)_{{\ka }},
\end{equation}
and write
\begin{equation}
\label{dobleco}
\frac{Det^{^{\prime }}(D)_{{\ka }}}{Det(i\not \! \partial
)_{{\ka }=0}}=\frac{Det(D+P_1)_{{\ka }}}{Det(i\not \! \partial
+P_0)_{{\ka }}}\ \frac{Det(i\not \! \partial  +P_0)_{{\ka }}}
{Det(i\not \!
\partial  )_{{\ka }=0}}.
\end{equation}

We can compute the first factor in the r.h.s. by taking the derivative 
\begin{equation}
\label{dalfa}
\dfrac \partial {\partial \alpha }\left[\ln Det{(D_{\alpha}
+P_{\alpha})_{{\ka
}}}\right]=Tr \left[ (\not \! \! A+ \partial_{\alpha}P_{\alpha} )
G_\alpha
\right],
\end{equation}
where $G_{\alpha}(x,y)$ is the Green's  function of the problem
\begin{eqnarray}
(D_\alpha +P_\alpha )\ G_{\alpha}(x,y)=\delta (x,y), \nn
\\
\left(
\begin{array}{cc}
{\cal P}_{\geq} & \, \,  {\cal P}_{\leq}
\end{array}
\right) G_{\alpha}(x,y)\vert _{r=R }=0.
\end{eqnarray}

Since $(D_{\alpha})_{{\ka}}$ is self-adjoint, $G_{\alpha}(x,y)$ has the
structure
\begin{equation}
\label{green2}
G(x,y)_{\alpha}=(1-P_\alpha )\ {\cal G}_{\alpha}(x,y)\
(1-P_\alpha )+\ P_\alpha,
\end{equation}
where ${\cal G}_{\alpha}(x,y)$ is the kernel of the right-inverse of
$D_\alpha $ on the orthogonal complement of Ker$(D_\alpha
)_{\ka}$\cite{spectral},
\begin{equation}
{\cal G}_{\alpha}(x,y)=\frac 1{2\pi i} \times
\left(
\begin{array}{cc}
0 & \frac{e^{\alpha [\phi (x)-\phi (y)]}}{X-Y}\left(
\frac XY\right) ^{k+1}
\\ \frac{e^{-\alpha [\phi (x)-\phi (y)]}}{X^{*}-Y^{*}}\left(
\frac{Y^{*}}{
X^{*}}\right) ^{k+1} & 0
\end{array}
\right) ,
\end{equation}
which, replaced in (\ref{green2}), allows to get
 $G(x,y)_{\alpha}$.

Being $P_\alpha$ an orthogonal proyector, 
\be
(P_\alpha)^2=P_\alpha, \qquad
\frac{\partial P_\alpha}{\partial \alpha} \left( 1 - P_\alpha\right) =
P_\alpha \frac{\partial P_\alpha}{\partial \alpha},
\ee
from (\ref{green2}) we get $Tr \left[ (\partial_\alpha P_\alpha )\,
G_\alpha \right]=0$. So, (\ref{dalfa}) reduces to the evaluation of $Tr
\left[ \not \! \! A G_\alpha \right]$. 

As usual, the kernel of the operator inside the trace is singular at the
diagonal, so we must introduce a regularization. We will employ a
point-splitting one where, following Schwinger \cite{schwinger-ps}, we
will introduce a phase factor in order to preserve gauge invariance. We
thus get,
\[
Tr \left[ \not  \! \! A G_\alpha \right]=
\]
\begin{equation}
{\rm  sym.\ lim.\ }_{\epsilon \rightarrow 0}\int_{r<R} d^2x \, \,
tr\left[ \not
\! \! A(x) G_\alpha(x,x+\epsilon) e^{i \alpha \epsilon \cdot A(x)}
\right] ,
\end{equation}
where by symmetric limit we mean half the sum of the lateral limits $\epsilon
\rightarrow 0^{\pm}$.

Performing the integral in $\alpha$ from 0 to 1 we get\cite{spectral}
\[
\ln
\left[
\frac{Det(D+P_1)_{{\ka }}}{Det(i\not \! \partial +P_0)_{{\ka }}}
\right] =-\frac
1{2\pi }\int _{r<R} d^2x\  {\phi ^{\prime }}^2
\]
\begin{equation}\label{a1}
 -2\ (k+1)\ \phi (R)\
+\sum\limits_{n=0}^k\
\ln \left[2(n+1)\frac{q_n(R;1)}{R^{2(n+1)}}\right]\ .
\end{equation}
Notice that when there are no zero modes $(k+1=0)$ only the first term in
the r.h.s. survives. 

\bigskip

In the following, we will obtain the second quotient of determinants in
Eq. (\ref{dobleco}) by computing explicitly the spectra of the free Dirac
operators and using a $\zeta$-function regularization. 

The eigenfunctions of ${(\ \not \!\!\!i\partial +P_0)_{{\ka }}}$ are of
the form
\begin{equation} 
\psi_n(r,\theta)=\left( \QATOP{\varphi_n (r,\theta )}
{\chi_n (r,\theta )}\right)
=\left( \QATOP{J_n(\vert \lambda \vert r)\
e^{in\theta }}{-i\frac{\vert \lambda \vert} { \lambda
}J_{n+1}(\vert \lambda \vert r)\ e^{i(n+1)\theta }}\right),
\label{autofun}
\end{equation}
and satisfy the boundary condition
\begin{equation}
{\cal P}_{\geq}\, \varphi_n (R,\theta ) =
 \frac 1{2\pi }\sum_{n\geq k+1}e^{in\theta }\left(
e^{in\theta
}, \varphi_n (R,\theta )\right) =0,
\end{equation}
\begin{equation}
{\cal P}_{\leq} \,\chi_n (R,\theta )
=\frac 1{2\pi }\sum_{n\leq k+1}e^{in\theta }\left( e^{in\theta
},\chi_n (R,\theta )\right) =0 .
\end{equation}
For $n\geq k+1$ the corresponding eigenvalues are $\lambda =\pm j_{n,l}/R$
( $j_{n,l}$ is the $l$-th zero of $J_n(z)$). Analogously, for $n\leq k,$
$\lambda =\pm j_{n+1,l}/R.$ Notice that $j_{-n,l}=j_{n,l} $, and that, for
$n=k+1$ the eigenvalues appear twice, once for an eigenfunction with
vanishing upper component at the boundary, and once for another one with
vanishing lower component. 

For $\Re (s)$ large enough, we can construct the $\zeta$-function of ${(\
\not \!\!\!i\partial +P_0)_{{\ka }}}$ as\cite{spectral}
\[
\displaystyle \zeta _{(i\not\partial  +P_0)_{{\ka }}}(s)=
\vert k+1\vert
+(1+e^{-i\pi s}) \times
\]
\begin{equation}
\label{zeta}
\left\{ \sum\limits_{n=-\infty }^{\infty }\sum\limits_{l=1}^\infty \
\left(
\dfrac{
j_{n,l}}R\right) ^{-s}+\sum\limits_{l=1}^\infty \ \left( \dfrac{j_{\vert
k+1\vert,l}}%
R\right) ^{-s}\right\} .
\end{equation}
The first term, $\vert k+1\vert $, is just the multiplicity of the
0-eigenvalue of $(i\not \! \partial )_{{\ka }}$. It is also interesting to
note that the double sum in the r.h.s. (which is independent of $k$)
corresponds to the $\zeta$-function of the Laplacian on a disk with
Dirichlet (local) boundary conditions, thus being analytic at $s=0$
\cite{seeley-trazas}. 

It is easy to verify that the analytic extension of the
second sum,
\begin{equation}
f_\nu (s)\equiv \sum_{l=1}^\infty \ (\ j_{\nu,l})^{-s}.
\end{equation} 
is regular at $s=0$. Then $\zeta _{(i\not\partial +P_0)_{{\ka }}}(s)$ is
regular at the origin. This is interesting since, as far us we know, the
regularity of the $\zeta$-function at the origin for non local boundary
conditions has not been established in general \cite{seeley-nl}. 

In the framework of this regularization, we thus get
\[
\ln \left[
\dfrac{Det(i\not \! \partial  +P_0)_{\ka }}{Det(i\not \!
\partial  )_{{\ka }
=0}}
\right] \equiv
-\dfrac d{ds}\left[ \zeta _{(i\not\partial  +P_0)_{{\ka }}}(s)-\zeta
_{(i\not\partial  )_{{\ka=0 }}}(s)\right] _{s=0} =
\]
\begin{equation}
\label{qdl}
-2\left[ f_{\vert k+1\vert}^{\prime }(0)-f_0^{\prime }(0)+
(\ln R-\frac{i\pi }
2)[f_{\vert k+1\vert}(0)-f_0(0)]\right] .
\end{equation}

Taking into account the asymptotic expansion for the zeros of Bessel
functions \cite{abram}, we obtain
\begin{equation}
\label{f0}f_\nu (0)=-\frac \nu 2-\frac 14,
\end{equation}
and
\begin{equation}
\label{f'0}f_\nu ^{\prime }(0)=-\frac 12\ln 2+\left( \frac{2\nu -1}4\right)
(\ln \pi -\gamma )-\sum\limits_{l=1}^\infty \ln \left[ \frac{\ j_{\nu ,l}}
{l\pi }\ e^{-\left( \frac{2\nu -1}{4\ l}\right) }\right] ,
\end{equation}
where $\gamma $ is Euler's constant. 

Finally, taking into account that we have
used a gauge invariant procedure, we can write
\begin{eqnarray}
\label{final}
\displaystyle \ln\left[
\frac{Det(D+P_1)_{{\ka }}}{Det(i\not \! \partial )_{{\ka }=0}}
\right] =-\frac
1{2\pi }\int_{r<R} d^2x\ A_\mu (\delta_{\mu \nu}-
\dfrac{\partial_{\mu}\partial_{\nu}}{\partial^2}) \ A_\nu 
\nn \\ \nn \\ 
-2\ (k+1)\ \phi(R)  \displaystyle +\sum\limits_{n=0}^k\
\ln\left[2(n+1)\frac{q_n(R;1)}{R^{2(n+1)}} \right] 
\\ \nn \\
 -\vert k+1\vert [\frac{i\pi }2-\gamma -\ln (\frac R\pi
)]+2\sum\limits_{l=1}^\infty \ln \left[ \frac{\ j_{\vert k+1\vert,l}}{\
j_{0,l}}\
e^{-\left( \frac{\vert k+1\vert}{2\ l}\right) }\right] . \nn
\end{eqnarray}
The first term is the integral on the disk of the same expression
appearing in the well-known result for the boundaryless case
\cite{gamboa}.

\subsection*{ Connection with the index theorem}

The variation of the determinant under global axial transformations
($\epsilon$ constant),
\[
 e^{-\gamma_{5} \epsilon }
(D+P_1)_{{\ka }}e^{-\gamma_{5} \epsilon }
=(D+e^{-\gamma_{5} \epsilon }P_1e^{-\gamma_{5} \epsilon })_{{\ka }} ,
\]
\begin{equation}
e^{-\gamma_{5} \epsilon }  (i\not \! \partial )_{{\ka }=0}e^{-\gamma_{5}
\epsilon }=(i\not \! \partial )_{{\ka }=0} ,
\end{equation}
is related to the {\it index} of the Dirac operator:
\begin{equation}
\label{dep}
\displaystyle \dfrac{\partial}{\partial \epsilon}\ln\left[
\frac{Det\left(e^{-\gamma_{5} \epsilon }(D+P_1)_{{\ka }}e^{-\gamma_{5}
\epsilon
}\right)}
{Det\left( e^{-\gamma_{5} \epsilon }(i\not \! \partial )_{{\ka
}=0}e^{-\gamma_{5} \epsilon }\right) }
\right]  =
 -2Tr \left[ \gamma _{5} P_1  \right]=-2(N_{+}-N_{-}),
\end{equation}
where  $N_{+(-)}$ is the number of positive(negative) chirality zero modes.

\bigskip

It can be verified that our strategy leads to the correct result for this
index. By following the same procedure that lead to Eq. (\ref{final}), we
can compute the quotient of determinants in the l.h.s of (\ref{dep}). In
fact, taking into account that the inverse of the transformed operator is
given by
\begin{equation}\label{green3}
G^{(\epsilon)}_{\alpha}(x,y)=(1-P_{\alpha} )\ {\cal G}_{\alpha}(x,y)\
(1-P_{\alpha} )+e^{\gamma_{5} \epsilon }\
P_{\alpha}\,  e^{\gamma_{5} \epsilon } ,
\end{equation}
the only difference appears in the first term
of the r.h.s. of (\ref{zeta}), where a factor $e^{\pm 2\epsilon s}$ arises.
Thus, after performing the $\epsilon$-derivative
\begin{equation}
\label{enemas}
N_{+}  -N_{-}= k+1,
\end{equation}
which agrees with our previous result for the number of zero modes.

\bigskip

The Atiyah-Patodi-Singer theorem relates the ${\rm index }(D)_{{\ka }}$
with the spectral asymmetry of the self-adjoint operator
\be
{\cal A}={\cal A}(R)=-\frac i R \ \partial _\theta +\ \partial _r \phi (R).
\ee
From the eigenvalues of ${\cal A}$, $a_n = \frac 1R(n-{\k})$, one defines
the $\eta$-function through the series
\begin{equation}
\eta_{({\cal A})}(s)= R^s \sum_{n \neq {\k }} {
sig(n-{\k } ) \, \,  \vert n-{\k } \vert^{-s}},
\end{equation}
convergent for $\Re (s) > 1$. The analytic extension of $\eta_{({\cal
A})}(s)$ to $s=0$ is given by \cite{spectral}
\begin{equation}
\eta_{({\cal A})}(0)=2({\k }-k)-1-{ h}({\cal A}),
\end{equation}
where $h({\cal A})=$ dim Ker$({\cal A})$.

Following the construction of APS in \cite{aps}, and taking into account
that $\varrho =\varrho(\theta) =-\ i\ e^{i\theta }$ in the present case,
we get
\begin{equation}
{\rm index}\,  D= {\k } + \frac{ \left[ 1- { h}({\cal A})-
\eta_{({\cal A})}(0)\right]}{2}=k+1,
\end{equation}
in agreement with (\ref{enemas}). The first term in the intermediate
expression is the well known contribution from the bulk \cite{gamboa}. The
second one is the boundary contribution of APS, shifted by 1/2. This
correction, due to the presence of the factor $\varrho$ in (\ref{op}), has
already been obtained in \cite{sitenko} with slightly different spectral
boundary conditions. 

\section*{Conclusions}

We have achieved the complete evaluation of the determinant of the Dirac
operator on a disk, in the presence of an axially symmetric flux, under
global boundary conditions of the type intoduced by Atiyah, Patodi and
Singer. 

To this end, we have proceeded in two steps: In the first place, we have
grown the gauge field while keeping the boundary condition fixed. This
calculation was possible thanks to the exact knowledge of the zero modes
and the Green's function (in the complement of the null space.) Here, a
gauge invariant point splitting regularization was employed. 

In the second step, we have explicitly obtained the eigenvalues of $
(i\not \!\partial +P_0)_{{\ka }}$. We have shown that the corresponding
$\zeta$-function is regular at the origin and we have evaluated the
quotient of the free Dirac operators for two different global boundary
conditions. 

We have verified that our complete result is in agreement with the APS
index theorem.


\begin{thebibliography}{10}

\bibitem{bordag1} M.~Bordag, B.~Geyer, K.~Kirsten, and E.~Elizalde.
\newblock Zeta function determinant of the laplace operator on the d-
dimensional ball. \newblock {\em hep-th}, (9505157), 1995. 

\bibitem{Dowker3:1995} J.S. Dowker and J.S. Apps. \newblock Functional
determinants on certain domains. \newblock {\em hep-th}, (9506204), 1995. 

\bibitem{Elizalde1:1996} E.~Elizalde, M.~Lygren, and D.V. Vasilevich.
\newblock Antisymmetric tensor fields on spheres: functional determinants
and non--local counterterms. \newblock {\em hep-th}, (9602113), 1996. 

\bibitem{kirsten-cognola} Klaus Kirsten and Guido Cognola. \newblock Heat
kernel coefficients and functional determinants for higher spin fields on
the ball. \newblock {\em Class. Quant. Grav.}, 13:633--644, 1996. 

\bibitem{Esposito1:1991} Peter~D. D'Eath and Giampiero Esposito. \newblock
Local boundary conditions for the {D}irac operator and one - loop quantum
cosmology. \newblock {\em Physical Review}, (D43):3234, 1991. 

\bibitem{Esposito2:1991} Peter~D. D'Eath and Giampiero Esposito. \newblock
Spectral boundary conditions in one - loop quantum cosmology. \newblock
{\em Physical Review}, (D44):1713, 1991. 

\bibitem{wipf} A.~Wipf and S.~Durr. \newblock {G}auge theories in a bag.
\newblock {\em Nuclear Physics}, B443:201 -- 232, 1995. 

\bibitem{bolsa-quiral} M.~De Francia, H.~Falomir, and E.~M. Santangelo.
\newblock {F}ree energy of a four dimensional chiral bag. \newblock {\em
Physical Review}, D45(6):2129--2139, 1992. 

\bibitem{mitdef} M.~De Francia. \newblock {F}ree energy for massless
confined fields. \newblock {\em Physical Review}, 50D:2908--2919, 1994. 

\bibitem{Esposito1:1997} Giampiero Esposito. \newblock Dirac {O}perator
and {S}pectral {G}eometry. \newblock {\em hep-th}, (9704016), 1997. 

\bibitem{mochado} H.~Falomir, R.~E.~Gamboa Sarav\'{\i}, M.~A. Muschietti,
E.~M. Santangelo, and J.~E. Solom\'{\i}n. \newblock {D}eterminants of
{D}irac operators with local boundary conditions. \newblock {\em Journal
of Mathematical Physics}, 37(11), 1996. 

\bibitem{bullito} H.~Falomir, R.~E.~Gamboa Sarav\'{\i}, M.~A. Muschietti,
E.~M. Santangelo, and J.~Solomin. \newblock On the relation between
determinants and {G}reen functions of elliptic operators with local
boundary conditions. \newblock {\em Bulletin des Sciences
Math\'ematiques}, in press, 1996. 

\bibitem{seeley-cb} R.~T. Seeley. \newblock {\em Am. J. Math.},
91:889--920, 1969. 

\bibitem{aps} M.~F. Atiyah, V.~K. Patodi, and I.~M. Singer. \newblock {\em
Math. Proc. Camb. Phil. Soc.}, 77:43, 1975. 

\bibitem{egh} P.~B.~Gilkey T.~Eguchi and A.~J. Hanson. \newblock {\em
Physics Reports}, 66:213, 1980. 

\bibitem{calderon} A.~P. Calder\'on. \newblock {\em {L}ectures notes on
pseuddifferential operators and elliptic boundary value problems, I}.
\newblock Publicaciones del I.A.M., Buenos Aires, 1976. 

\bibitem{Stein} E.~M. Stein. \newblock {\em {S}ingular {I}ntegrals and
{D}ifferentiability {P}roperties of {F}unctions}. \newblock Princeton
University Press, Princeton, New Jersey, 1970. 

\bibitem{hormander} L.~H{\"o}rmander. \newblock {\em {T}he {A}nalysis of
{L}inear {P}artial {D}ifferential {O}perators {III},
{P}seudo-{D}ifferential {O}perators}. \newblock Springer-Verlag, Berlin
Heidelberg, 1985. 

\bibitem{booss-b} B.~Booss and D.~Bleecker. \newblock {\em {T}opology and
{A}nalysis. {T}he {A}tiyah-{S}inger {I}ndex {F}ormula and
{G}auge-{T}heoretic {P}hysics}. \newblock Springer - Verlag, New York,
1985. 

\bibitem{booss-w} D.~Booss and K.P. Wojciechowski. \newblock {\em
{E}lliptic {B}oundary {P}roblems for {D}irac {O}perators}. \newblock
Birkh{\"a}user, Boston, 1993. 

\bibitem{seeley-trazas} R.~T. Seeley. \newblock {A}nalytic extension of
trace associated with elliptic boundary problem. \newblock {\em American
Journal of Mathematics}, 91:963--983, 1969. 

\bibitem{seeley-nl} G.~Grubb and R.~T. Seeley. \newblock {Z}eta and eta
functions for {A}tiyah - {P}atodi - {S}inger operators. \newblock Univ.~of
Copenhagen, Math.~Inst., preprint Nro.~11, 1994. 

\bibitem{spectral} H.~Falomir, R.~E.~Gamboa Sarav\'{\i}, and E.~M.
Santangelo. \newblock {D}irac operator on a disk with global boundary
conditions. \newblock {\em hep-th/9609194}, 1996. 

\bibitem{schwinger-ps} J.~Schwinger. \newblock {\em Physical Review},
82:664, 1951. 

\bibitem{abram} M.~Abramowitz and I.~Stegun. \newblock {\em {H}andbook of
{M}athematical {F}unctions}. \newblock Dover Publications, 1970. 

\bibitem{gamboa} R.~E.~Gamboa~Sarav\'\i \, M.~A. Muschietti, F.~A.
Schaposnik, and J.~E. Solomin. \newblock {\em Annals of Physics}, 157:360,
1984. 

\bibitem{sitenko} A.~V. Mishchenko and Yu.~A. Sitenko. \newblock {\em Ann.
Phys.}, 218:199, 1992. 

\end{thebibliography}

\end{document}